\documentclass [12pt]{article}
\usepackage[dvips]{color}
\usepackage{epsfig}
\usepackage{color}
\def\beq{\begin{equation}}
\def\eeq{\end{equation}}
\def\beqa{\begin{eqnarray}}
\def\eeqa{\end{eqnarray}}
\def\d{{\rm d}}
\def\ttimes{{\scriptstyle \times}}
\def\rc{{\mbox{\tiny \rm c}}}
\def\rmm{{\mbox{\tiny \rm m}}}
\def\rU{{\mbox{\tiny \rm U}}}
\def\rf{{\mbox{\tiny \rm f}}}
\def\rL{{\mbox{\tiny \rm L}}}
\def\rint{{\mbox{\tiny \rm int}}}

\begin{document}
\baselineskip0.6cm plus 1pt minus 1pt
\tolerance=1500
{\vspace*{-1cm}
\flushleft{Artigos gerais}
\vspace*{1cm}}
\begin{center}
{{\Large\bf  A 4-vector formalism for classical mechanics} \\
{(Um formalismo de tetravetores para a mec\^anica cl\'assica)}}

\vskip0.8cm
{ J. G\"u\'emez$^{a}$,
M. Fiolhais$^{b,}$\footnote{tmanuel@teor.fis.uc.pt}}
\vskip0.3cm
{\it $^a$ Departamento de F\'{\i}sica Aplicada}\\ {\it Universidad de
Cantabria} \\ {\it E-39005 Santander, Spain} \\
\vskip0.2cm
{\it $^b$ Departamento de F\'\i sica and Centro de
F\'\i sica Computacional}
\\ {\it Universidade de Coimbra}
\\ {\it P-3004-516 Coimbra, Portugal}
\vskip0.2cm
\end{center}

\begin{abstract}
We present a matrix formalism, inspired by the Minkowski four-vectors of special relativity, useful to solve classical physics problems
related to both  mechanics and thermodynamics.
The formalism turns out to be convenient to deal with exercises involving non-conservative forces and production or destruction of mechanical energy.
On the other hand, it provides a framework to treat straightforwardly changes of inertial reference frames, since it embodies the Principle of Relativity.
We apply the formalism to a few cases to better show how it works.


\vskip0.5cm

Apresentamos um formalismo matricial, inspirado nos tetravetores de Minkowski da teoria da relatividade, \'util para resolver problemas de f\'\i sica cl\'assica relacionados com
mec\^anica e termodin\^amica.
O formalismo \'e especialmente conveniente quando se aplica a exerc\'\i cios que envolvam for\c cas n\~ao-conservativas e produ\c c\~ao ou destrui\c c\~ao de energia mec\^anica. Por outro lado, o formalismo permite lidar de forma direta com mudan\c cas de referencial de in\'ercia pois ele incorpora o princ\'\i pio da relatividade.
Para melhor se apreciar o seu funcionamento, aplicamos o formalismo a alguns casos concretos.

{\noindent {\bf Palavras-chave:} mec\^anica cl\'assica, termodin\^amica, tetravetores de Minkowski, ensino da f\'\i sica}
\end{abstract}

\section{Introduction}

In the literature, it is quite often to find papers motivated by the bridge between classical mechanics and the first law of thermodynamics \cite{lehrman73,arons89,leff93,guemez13}
and this work is one contribution towards that objective, by means of the development of a new matrix formalism.

Classical mechanics is a powerful theory within its limit of applicability.
On the other hand, thermodynamics is a quite general theory formulated without any hypothesis
on the nature and constitution of the system.
Both are very well established and they are part of any
physics curriculum. Each one is based on a number of principles that were directly motivated by the observation of natural phenomena. Isaac Newton established the
basis and the principles of mechanics, in the 17th century, whereas the thermodynamics principles were established essentially in the course of the 19th century as the result of the
efforts of a number of scientists (James Joule,  Lord Kelvin,  Ludwick Boltzmann,  Rudolf Clausius, Sadi Carnot, Walther Nernst, among many others).
Interestingly enough, in the university physics curricula, mechanics and thermodynamics almost do not intercept, although this is not the case in everyday life where
both are connected and very tightly bound: the automobile is probably the most common example \cite{guemez13b} but there is a myriad of other examples.

In this paper we present a  matrix formalism that embodies and tackles simultaneously mechanics and  the first law of thermodynamics.
It is
inspired by the Minkowski four-vector formulation of special relativity, which is a most elegant, successful  and powerful
formalism \cite{freund08}.
The main idea  is to capture the essential aspects of the formulation of the theory of relativity with four-vectors,
and incorporate those aspects in a similar formalism suitable for classical physics.
To this end  we define classical 4-vectors to describe the  system (space-time coordinates, momentum-kinetic energy, impulse-work, etc.) and represent the interactions, i.e. the  forces, through
4$\ttimes 4$ matrices. These matrices are inspired and emulate the electromagnetic field tensor in special relativity.
 In summary, we combine basic laws of mechanics and thermodynamics in a matrix formalism, whose essence lies in the special relativity, that we find
particularly useful in solving classical mechanics problems which require both a dynamical and an energetic analysis.

We apply the formalism to two concrete situations: a person jumping vertically; and a block pulled by a force, sliding with friction on top of a horizontal plane. The  dissipative work of the friction force, following other authors \cite{besson01},
 is considered as heat in the energy balance equation. Regarding the mechanical description, this  choice is immaterial, since the pertinent quantity is the force itself (assumed to be constant) times the displacement of the centre-of-mass (the so-called pseudo-work \cite{penchina78}).
However, for the energetic description we show that treating the dissipative work simply as heat provides a clearer description  as far as the first law of thermodynamics is concerned.
   Such perspective, which is possible because there is an immediate and total conversion of all dissipative work into heat is, of course, not unique  but our choice is better suited for the thermodynamical analysis.

This paper is organized as follows.
In section \ref{sec:clasmeter} we present the essentials of classical mechanics and thermodynamics that are pertinent for the derivation of
our  formalism. This section may seem rather pedestrian, but it already
tackles some subtle points of  the first law of thermodynamics on their relationship with mechanics.
In section \ref{sec:relwfvec}  we review the four-vector Minkowski formalism. Again, though it might seem a pedestrian formulation, we try to emphasize those aspects that are more subtle  and usually not
explicitly mentioned in textbooks, such as the existence of an impulse-work / momentum-energy equation.
In section \ref{sec:fvfcmt} we derive the new matrix formalism for classical physics, underlying its robustness and capabilities, which also include the automatic implementation of the relativity principle.
In section \ref{sec:tformwor}  we treat two problems of mechanics, which also involve  production and destruction of mechanical energy, to show the formalism at work and to
 illustrate both its elegance and powerfulness. Section \ref{sec:conclu}  is devoted to the conclusions.

\section{Classical mechanics and  the first law of thermodynamics}
\label{sec:clasmeter}

\subsection{Newton's second law}
\label{ssec:newslaw}

For a system of classical particles, of total {\em constant} mass $M$, the Newton's second law states that
the resultant of the external forces, ${\vec F}_{\rm ext}$, is equal to that mass times the centre-of-mass acceleration, or equivalently
 \beq
 M \d {\vec v}_{\rm cm}={\vec F}_{\rm ext}\  \d t  \ ,\label{newtonpt1}
\eeq
where $v_{\rm cm}$  stands for the  centre-of-mass velocity.
This is an infinitesimal form of Newton's fundamental law. Its integral form is $\Delta \vec p_{\rm cm}= \vec I $, where
$ \vec I = \int  {\vec F}_{\rm ext}\  \d t $ is the impulse of the resultant of the external forces, and states that this external impulse  in a given time interval is equal to the
variation, in the same time interval, of the system centre-of-mass linear momentum, $\vec p_{\rm cm}=M \vec v_{\rm cm}$.

The fundamental equation (\ref{newtonpt1}) still allows us to conclude that
the infinitesimal variation of the kinetic energy of the center-of-mass equals the so called
``pseudo-work" \cite{penchina78,sherwood83},
which is the dot product of the external resultant force applied to the system by the infinitesimal displacement of the
center-of-mass:
\beq
 {1\over 2} M \d v_{\rm cm}^2={\vec F}_{\rm ext} \cdot \d {\vec r}_{\rm cm}\, . \label{newtonptx}
\eeq
Actually, after scalar-multiplying both sides of (\ref{newtonpt1}) by $\vec v_{\rm cm}$ one readily arrives to (\ref{newtonptx}), the equivalence of the two equations, (\ref{newtonpt1}) and (\ref{newtonptx}),
being a consequence of $\d (v_{\rm cm}^2/2) / \d v_{\rm cm} =v_{\rm cm} $.
Equation~(\ref{newtonptx}) can still be written as
\beq
\d K_{\rm cm} = {\vec F}_{\rm ext} \cdot \d {\vec r}_{\rm cm} \, , \label{newtonpt2}
\eeq
where $K_{\rm cm}={1 \over 2} M v^2_{\rm cm}$ is the centre-of-mass kinetic energy.
The integral form of  equation  (\ref{newtonpt2})  is $\Delta K_{\rm cm}={W_{\rm ps}}$, where the left-hand side is the variation of the centre-of-mass kinetic energy
and the right-hand side is the ``pseudo-work" of the resultant external force, ${W_{\rm ps}}=\int {\vec F}_{\rm ext} \cdot \d {\vec r}_{\rm cm}$ \cite{mallin92}.
It is worthwhile to make  clear the distinction between work and pseudo-work: in the former, one considers the displacement of the force itself, whereas in the
latter, corresponding to the present case,  it is the displacement of the centre-of-mass that matters. Hence, the energy-like formulation of Newton's second law uses the pseudo-work and not the work of the resultant force a point that, surprisingly, is not much emphasized in textbooks.
Of course, there are circumstances for which both  work and pseudo-work are equal (if the displacement of the force is the displacement of the centre-of-mass) but this is not, in general, the case for a system of particles.

The above equations, though expressing the same fundamental law, do it in two different forms: (\ref{newtonpt1}) is a vector equation, and (\ref{newtonpt2}) is a scalar one.
The information that can be extracted is, sometimes, complementary, because they use different sets of physical magnitudes. However, they express the same physical law --- Newton's second law.
The first one uses the impulse and the linear momentum, and the second one
the pseudo-work and the kinetic energy. It is appropriate to note that equivalent descriptions of a physical system in terms of scalar quantities or of vector ones also happens in electrostatics, when the
scalar potential is used or, equivalently, when the electric field is used in the description of an electric charge system in static equilibrium.

\subsection{The first law of thermodynamics}

The previous equations are always valid and pertinent to describe the motion of the centre-of-mass of a system. However they do not
provide a full  physical description in general situations. To this purpose one still needs the first law of thermodynamics
which brings about new information. The mechanical system itself possesses an internal energy, $U$, whose variation may result from energetic
transfers to and from the system, either as work, $W$, or as heat, $Q$. It may also vary due to processes inside the system itself, e.g. when chemical reactions take place in the system. The pseudo-work does not play any role here: it is the real work that enters the energy balance expressed by the first law of thermodynamics.

In general, and for any system, the infinitesimal internal energy
variation, $\d U$, may receive contributions from the variation of the internal kinetic energy, $\d K_{\rint}$ (including rotational kinetic energy or kinetic energy with respect to the centre-of-mass),
from any internal work, $\d W_{\rm int}$ (i.e. work performed by the internal forces) \cite{mallin92},
from the internal energy variations related to temperature variations, $ Mc\d T$ ($c$ is the specific heat),  from the internal energy variations
related to chemical reactions \cite{atkins10}, etc.
The internal energy variation caused by a variation of the internal kinetic energy is particularly important
for mechanical systems that undergo processes lying in the scope of thermodynamics \cite{guemez13}.

The first law of thermodynamics is a statement on energy conservation and,
for a general process on a macroscopic system, whose analysis needs to combine mechanics and thermodynamics \cite{jewett08v},
it can be expressed, in infinitesimal form, as \cite{besson01,erlichson84}
\beq
\d {K}_{\rm cm} + \d { U} = \sum_j { {\vec F}_{{\rm ext}, j}}\cdot \ \d {\vec r}_j  \ + \ \delta { Q}.
\label{totale1}
\eeq
Here, the infinitesimal heat is denoted by $\delta Q$ since it is not an exact differential, contrary to $\d U$, $\d K_{\rm cm}$ or $\d {\vec r}_j$ that are scalar or vector exact differentials.
Each term in the sum over $j$ on the right-hand side of equation~(\ref{totale1})
is work associated with each {\em external} force ${\vec F}_{{\rm ext}, j}$, and  $\d {\vec r}_j$
is the infinitesimal displacement of that  force itself (and not, anymore, the displacement of the centre-of-mass). Therefore, $\delta  {W}_j= {\vec F}_{{\rm ext}, j} \cdot \d {\vec r}_j$
is real work (not pseudo-work).
At least in principle, if the right-hand side of equation (\ref{totale1}) vanishes, the internal energy may transform into kinetic energy, and vice-versa, if the process is compatible with  the Newton's second law (and with the second law of thermodynamics).
 Equation  (\ref{totale1}) also makes it clear that any external energetic interaction may serve to change the internal energy of the system or to change its centre-of-mass kinetic energy. The potential energy of the system as a whole may change if some of the external forces are conservative.
In that case, the work of those external forces, included  on the right-hand side of (\ref{totale1}), is equivalent to a corresponding potential energy variation on the left-hand side (remember that the variation of the potential energy associated with a conservative force, $\vec F_{\rm c}$, is defined through $\Delta E_{\rm p}= -\int {\vec F}_{\rm c}\cdot \ \d {\vec r}$).

The integral form of equation (\ref{totale1}) is expressed by $\Delta K_{\rm cm}+ \Delta U= W_{\rm ext} + Q$ or, equivalently, given by  $\Delta K_{\rm cm}+ \Delta E_{\rm p}+\Delta U= W + Q$, where the total potential  energy variation is already explicitly taken on the left-hand side and, therefore, on the right-hand side one has to exclude that part from the external work, the remaining part being now simply denoted by $W$. In a reductionist sense, $W$ could appropriately be called the thermodynamical work. In the most common thermodynamic examples there is no variation of the centre-of-mass kinetic or potential energy, i.e.   $\Delta K_{\rm cm}= \Delta E_{\rm p}=0$ and this is why the first law of thermodynamics  appears, in thermodynamical textbooks, formalized just as $\Delta U=W+Q$. We stress that, in this form, the right-hand side of the equation refers to energetic transfers to/from the system surroundings, therefore it is an {\em external} energy transfer. It is worth noting that  equation~(\ref{totale1}) allows for a  physical description of a thermal engine that  moves by itself (even moves up), for instance, the Stephenson's  Rocket \cite{stephen}, or  an accelerating  car \cite{guemez13b} or a walking person or toy \cite{guemez13c}. That would  not be possible using the first law as given by the well known expression $\d U = \delta W + \delta Q$.

Equations (\ref{newtonpt2}) and (\ref{totale1}) both refer to energy balances in processes but they express two quite different fundamental physical laws. The former, the so-called centre-of-mass energy equation,
is an alternative  way of stating Newton's second law, whereas the latter expresses the first law of thermodynamics. They are both simultaneously general and always valid,
therefore each one provides new information with respect to the other. Of course, if the problem is out of the scope of thermodynamics  (e.g. a mechanical system without friction), the two equations
are then equivalent or, in other words, they become the same equation. This happens when there is no internal energy variation, no heat transfers and when the displacement
of the forces equals the displacement of the centre-of-mass. The sum on the right-hand side of equation~(\ref{totale1}),
$   \sum_j { {\vec F}_{{\rm ext}, j}}\cdot \ \d {\vec r}_j$, then simply becomes  $\vec F_{\rm ext}\cdot \ \d {\vec r}_{\rm cm}$, where $\vec F_{\rm ext}= \sum_j { {\vec F}_{{\rm ext}, j}}$ is the resultant of the external forces, and (\ref{totale1}) is equal to (\ref{newtonpt2}).

\subsection{The Principle of Relativity}

The principle of relativity only played a minor role in the development of classical mechanics \cite{kleppner10}.
However it is a very important concept and all above equations must obey that principle, so they are valid in any inertial frame. There are physical quantities that do not change  upon  an inertial frame transformation, such as the forces, the mass, the time interval, the variation of internal energy and the heat. Other quantities do change, such as the variation of the centre-of-mass kinetic energy or the spatial displacement of a force that vary  from one inertial reference frame to another.
But equations (\ref{newtonpt1}) -- (\ref{totale1}) are Galilean invariant.
This means that  the amount of information provided by these equations will be {\em exactly} the same, irrespective of the chosen inertial frame --- there are no privileged inertial frames.

\section{Relativity with four-vectors}
\label{sec:relwfvec}

A Minkowski four-vector embodies, in the same entity, a ``space-like" (three-vector) part and a ``time-like" (scalar) part:
This is precisely the nature of, respectively,
equation (\ref{newtonpt1}), on the one hand, and of equations (\ref{newtonpt2}) or (\ref{totale1}),  on the other hand. This correspondence was actually a motivation
for developing a formalism, inspired in the  formulation of special relativity with Minkowski four-vectors, applicable in classical physics. That formalism will be worked-out in the
next section.

In this section we review those aspects of the relativistic formalism that are essential to derive the corresponding classical one \cite{guemez11b}.
In the Minkowski's  formulation of relativity, a basic physical magnitude is usually expressed as a four-vector.
For instance,  the  momentum-energy, $p^\mu$,  or the space-time, $r^\mu$, are (contravariant) four-vectors
given by the column matrices:
\beq
p^\mu = \left(\begin{array}{c} p_x\\p_y\\p_z \\  E/c \end{array}\right) \, ; \ \ \ \
r^\mu = \left(\begin{array}{c}  x\\ y\\ z \\  c  t \end{array}\right)\, .
\label{prp}
\eeq
On the other hand, the interaction of a particle with a field is expressed by means of a $4\times 4$-tensor.
For example,  the   electromagnetic field tensor ${\rm F}_\nu^{\mu}$, associated just with the electric force, $\vec F = q \vec E$, where $q$ is the electric charge and   $\vec E = (E_x, E_y, E_z)$ the electric field,  is such that:
\beq
q \, {\rm F}_\nu^\mu  = \left(\begin{array}{cccc}0 & 0 & 0 &  qE_x \\0 & 0 & 0 & qE_y \\0 & 0 & 0 & qE_z \\  qE_x & qE_y & qE_z & 0\end{array}\right)
 = \left(\begin{array}{cccc}0 & 0 & 0 &  F_x \\0 & 0 & 0 & F_y \\0 & 0 & 0 & F_z \\  F_x & F_y & F_z & 0\end{array}\right)\, .
\label{tensor}
\eeq
The matrix in (\ref{tensor}) refers to a charged particle observed in a reference frame in which there is only an electric field. In another inertial reference frame, the electromagnetic field tensor would acquire  other components. However, the matrix as given by (\ref{tensor}) is the pertinent one for the derivation of the classical formalism in the next section.
In relativity, the equations are usually better
expressed as products of $4\times 4$ matrices and four-vectors. For a particle with inertia ${ M}$, moving with velocity
${\vec v} = (v_x, v_y, v_z)$ in reference S,  its momentum-energy can also be expressed by  $E^\mu = c p^\mu$  explicitly given by
 \cite{guemez11c}:
\beq
E^\mu = \left(\begin{array}{c} c \gamma  {M} v_x\\c \gamma  { M} v_y \\c \gamma  { M} v_z \\ \gamma  { M} c^2 \end{array}\right)
\eeq
where $\gamma  = \left[1 - \beta^2 \right]^{-1/2}$, $\beta =v/c$ are  the usual relativistic coefficients. In the case of an electric charge moving in an external background electric field its relativistic description
is formally very similar  to the classical treatment described in section 2.1.
A four-vector infinitesimal impulse-work, $\d {W}^\mu$, can be defined by
\cite{guemez10b}
\beq
\d {W}^\mu = \left(\begin{array}{c} c \, \d {I}_x\\c \, \d {I}_y\\c \, \d {I}_z \\ \d W\end{array}\right) = \left(\begin{array}{cccc}0 & 0 & 0 &  F_x \\0 & 0 & 0 & F_y \\0 & 0 & 0 & F_z \\  F_x & F_y & F_z & 0\end{array}\right)\left(\begin{array}{c}  \d x\\ \d y\\ \d z \\  c \,  \d t \end{array}\right)\, .
\label{fgty}
\eeq
In the absence of magnetic fields, i.e. for a charged particle moving in an external electric field,
the motion of the body is governed by the four-vector fundamental equation \cite{guemez10b} which is the relativistic counterpart of the centre-of-mass equation (\ref{newtonpt2})
\beq
\d {E}^\mu = \d {W}^\mu \, .
\label{eqrelat}
\eeq
This `momentum-energy / impulse-work equation' is completely equivalent to the more familiar equation $F^\mu=\d p^\mu / \d \tau$, where $F^\mu$ is the four-force, $\d \tau = \gamma^{-1} \d t$ is a proper time and, hence, $cF^\mu \d \tau=\d W^\mu$, which is equation (\ref{eqrelat}). Explicitly, that equation reads as
\beq
\left(\begin{array}{c} c  {M} \ \d \left[\gamma (v) v_x\right]\\c {M} \ \d \left[\gamma (v) v_y\right]\\c {M} \ \d \left[\gamma (v) v_z\right] \\ { M} \ c^2 \ \d \left[\gamma (v) \right] \end{array}\right) = \left(\begin{array}{c}  cF_x \d t\\ cF_y \d t\\ c F_z \d t \\  F_x\d x + F_y \d y + F_z \d z \end{array}\right)\, .
\label{explicit45}
\eeq
The set of the first three equations can be regarded as the relativistic vector Newton's second law (corresponding to equation (\ref{newtonpt1}) in classical mechanics) and the fourth equation as the relativistic
centre-of-mass equation (corresponding to equation (\ref{newtonpt2}) in classical mechanics).
It is worth noting that this relativistic centre-of-mass equation can be obtained  from the relativistic Newton's second law, and vice-versa, by using $\d (\gamma  c^2) / \d (\gamma v) = v$ \cite{tsamparlis10}.
This is the relativistic counterpart of the derivation that leads, in classical mechanics,  from (\ref{newtonpt1}) to (\ref{newtonptx}).
Hence, in a sense, there is some redundancy in the information provided by the set of four components of the four-vector equation~(\ref{explicit45}) of the same level of the existing redundancy in equations
(\ref{newtonpt1}) and (\ref{newtonpt2}).

If the motion of the body is now described in reference frame S$'$, moving with velocity $V$ along the common $xx'$ axes with respect to S (standard configuration),
the four-vector equation (\ref{eqrelat}) is readily transformed to S$'$, and this is a
major advantage of the formalism:
\beq
\d E'^\mu = \d W'^\mu\, .
\label{eqrelat2}
\eeq
The primed components are obtained from the unprimed ones by means of the Lorentz transformation matrix (asynchronous formulation)
 \cite{cavalleri69}
\beq
{ \Lambda}_\nu^\mu  (V) =\left(\begin{array}{cccc} \gamma   & 0 & 0 & -\beta \gamma  \\0 & 1 & 0 & 0 \\0 & 0 & 1 & 0 \\-\beta  \gamma  & 0 & 0 & \gamma\end{array}\right)\, .
\label{eq-1}
\eeq
In particular, $  \Lambda_\nu^\mu  (V)  \left( \d E^\nu = \d W^\nu\right) \, \rightarrow \, \d E'^\mu = \d W'^\mu\, .
$
In general, four-vectors, such as $E^\mu$, and tensors, such as ${\rm F}^\mu_\nu$, transform from S to S$'$ according to
\beq
{E'}^\mu = { \Lambda}_\nu^\mu  \, \,  E^\nu   \, , \ \ \ \ \
{\rm F'}_\nu^\mu  = { \Lambda}_\alpha^\mu  (V) \,  {\rm F}_\beta^\alpha \,  { \Lambda}^\beta _\nu (-V)\, ,
\eeq
where  ${ \Lambda} (-V) = {{ \Lambda}^{-1}} (V)$.

The formalism implies that the Principle of Relativity is automatically fulfilled, i.e. the four-vector equations
exhibit the same form in  S and in S$'$ inertial reference frames.
As a    consequence, the four equations provided by (\ref{eqrelat}) in one reference frame can always be expressed
as linear combinations of the four equations  provided by (\ref{eqrelat2}) in the other reference frame, and vice-versa.

\section{ Formalism with 4-vectors for classical mechanics}
\label{sec:fvfcmt}

Now, the question is whether a formalism for classical physics, which retains the characteristics of the Einstein-Minkowski theory of relativity,
might be derived. Apparently not, since the same velocity can never be assigned to the same  phenomenon in two different inertial reference frames.
However, let us assume that there is some limiting velocity, $v_\rL$, in classical mechanics --- the same in all inertial
reference frames.
The assumption helps, as we shall see, in defining classical 4-vectors  and set-up matrix equations  based on the equations presented in section~\ref{sec:clasmeter}.
Typical velocities in classical physics, including the relative velocities of moving inertial frames, will be always assumed to be very small compared with $v_\rL$, so that
the limit  $v_\rL \rightarrow \infty$ is supposed or explicitly taken at some stage of the derivations. The new formalism, to be derived below, is
useful to discuss classical physics problems, keeping the advantages of the relativistic formalism. In particular, the Galilean principle of
relativity is automatically incorporated in such formalism. On the other hand, by including  the first law of thermodynamics, it allows for an integrated vision and treatment of those problems involving thermal effects and
variations of mechanical energy.

To obtain a non-relativistic reduction of the four-vectors and of the  transformation matrix presented in the previous section, we replace the speed of light by the above mentioned velocity parameter, $c\rightarrow v_\rL$,
which, as already mentioned, is  much larger than any velocity in the classical system. The role of $v_\rL$ is similar to the role played by $c$ in relativity in the sense of providing components of the 4-vectors that  are dimensionally homogeneous. As we shall see, its explicit value is never needed (hence, we could even keep $c$ instead of introducing a new quantity  $v_\rL$).
The classical limits for the relativistic space-time components are simply $\vec r  \rightarrow  \vec r  \,$ and $\ c  t \rightarrow v_\rL  t$. For the momentum-energy components   they are (see equations (\ref{prp}))
\beq
c \vec p \rightarrow v_\rL \, M \vec v \, ; \ \ \ E \rightarrow {1\over2} M v^2 + M \, v_\rL^2+U\, .
\label{eqwaes}
\eeq
In (\ref{eqwaes}) there is a rest-energy like term, $Mv_\rL^2$,
as in  special relativity  \cite{hecht08}
that will be discussed later, and $U$,  the internal energy, is already included, anticipating that the classical system may undergo internal energy variations.
Besides the classical 4-vector momentum-kinetic energy  (now preferably represented by ${\cal K}$, whose components are expressed in units of energy) we also define a 4-vector internal energy
(related to the momentum-energy in the rest frame), ${\cal U}$. Later on in this section we shall discuss in more detail this 4-vector ${\cal U}$.

The two 4-vectors, ${\cal K}$ and ${\cal U}$,  and the 4-vector position, ${\cal R}$ (all 4-vectors are denoted by calligraphic letters) are
column matrices
given by:
\beq
{\cal K} = \left(\begin{array}{c} v_\rL M\, v_x  \\ v_\rL M\, v_y \\ v_\rL M\, v_z \\ {1\over 2} M \, v^2 \end{array}\right) \, ,\ \
{\cal U} = \left(\begin{array}{c} 0  \\ 0 \\ 0 \\  M v_\rL^2 + U \end{array}\right) \, ,\ \
 {\cal R} = \left(\begin{array}{c}   x \\   y \\ z \\  v_\rL  t \end{array}\right) \, , \label{eq-2}
\eeq
where  contravariant (or covariant) indexes have been omitted  since they play no relevant role. By including the rest energy $Mv_\rL^2$ in the 4-vector internal energy, the components of the classical 4-vector $\cal K$ become simply the linear momentum (times a constant factor $v_\rL$) and the kinetic energy.

On the other hand, the classical reduction of the Lorentz transformation matrix, equation (\ref{eq-1}),
keeping only terms up to order $V /v_\rL$, is
\beq
 \Lambda_\nu^\mu (V)  \, \rightarrow \, {\Upsilon} =\left(\begin{array}{cccc} 1 & 0 & 0 & -\beta (V) \\0 & 1 & 0 & 0 \\0 & 0 & 1 & 0 \\-\beta (V) & 0 & 0 &   1
\end{array}\right)
\label{upsilon}
\eeq
now with the following new  definition: $ \beta (V) = {V}/{v_\rL}$.
The matrix $\Upsilon$  is a symmetric one, its inverse is $\Upsilon^{-1}(V)=\Upsilon(-V)$ and its generalization for translations in any direction is straightforward,  with the property  $\Upsilon(\vec V_1+\vec V_2)=\Upsilon(\vec V_1)\Upsilon(\vec V_2)$.
One should note that this is not exactly the usual Galilean matrix transformation, since applying $\Upsilon$ to the 4-vector position leads to
\beq
\left( \begin{array}{c} x' \\ y' \\ z' \\ v_\rL t' \end{array}\right) = \left(\begin{array}{cccc} 1& 0 & 0 & -\beta \\0 & 1 & 0 & 0 \\0 & 0 & 1 & 0 \\-\beta & 0 & 0 & 1
\end{array}\right)\left( \begin{array}{c} x \\ y \\ z \\ v_\rL t \end{array}\right) = \left( \begin{array}{c}  x -\frac{V}{v_\rL} \, v_\rL t \\ y \\ z \\
 v_\rL t - \frac{V}{v_\rL} x  \end{array}\right)\, .
\label{t677}
\eeq
 If we take the limit $v_\rL \rightarrow \infty$, the Galilean transformation is recovered:
\beq
\left( \begin{array}{c} x' \\ y' \\ z' \\ v_\rL t' \end{array}\right)=
  \left( \begin{array}{c}  x - V t \\ y \\ z \\
 v_\rL t  \end{array}\right)\, . \label{t678}
\eeq
Up  to leading order in $v_\rL$, the time is absolute, i.e.  $t=t'$ [see (\ref{t677})], as stated by the Galilean transformation (\ref{t678}). Both the velocity 4-vector, which
is the time derivative of the position 4-vector, and its transformed 4-vector, are now readily obtained:
\beq
\dot{\cal R}= \left( \begin{array}{c} v_x \\ v_y \\ v_z \\ v_\rL  \end{array}\right); \ \ \ \mbox{and}
\ \
\dot{\cal R}' = \Upsilon  \dot{\cal R}=
  \left( \begin{array}{c}  v_x - V  \\ v_y \\ v_z \\v_\rL  \end{array}\right)
\eeq
The acceleration 4-vectors are
identical in S and S$'$: $\ddot{\cal R}'=\ddot{\cal R}$.

\subsection{Centre-of-mass equation}

Motivated by the form of the electric components of the electromagnetic field tensor [see (\ref{tensor})]
we associate  to each force $\vec F = (F_x, F_y, F_z)$ the following $4\times 4$ matrix:
\beq
{\bf F} = \left(\begin{array}{cccc}0 & 0 & 0 &  F_x \\0 & 0 & 0 & F_y \\0 & 0 & 0 & F_z \\  F_x & F_y & F_z & 0\end{array}\right) \label{18m}
\eeq
(we  denote the `classical' $4\times 4$ matrices representing forces by bold letters).

Let us now consider the following
classical matrix equation referring to the centre-of-mass, which can be regarded as the classical counterpart of the relativistic equation (\ref{eqrelat}) [that can also be written as $\d E^\mu = F^\mu \, \, (c \, \d \tau)$:
\beq
\d {\cal K}_{\, \rm cm} = {\bf F}_{\rm ext}\, \, \d {\cal R}_{\, \rm cm}  \label{eq-3}
\eeq
where ${\bf F}_{\rm ext}$  is a $4\times 4$ matrix representing the resultant of the {\em external} forces:  ${\bf F}_{\rm ext}= \sum_j{\bf F}_{{\rm ext},j}$.
The 4-vectors in equation (\ref{eq-3}) are explicitly given by [see equation (\ref{eq-2})]
\beq
\d  {\cal K}_{\, \rm cm} = \left(\begin{array}{c} v_\rL M\, \d v_{{\rm cm},x}  \\ v_\rL M\, \d v_{{\rm cm},y} \\ v_\rL M\, \d v_{{\rm cm},z} \\  M \, v_{\rm cm} \d v_{\rm cm}  \end{array}\right) \ ; \ \ \
\d  {\cal R}_{\, \rm cm} = \left(\begin{array}{c} \d  x_{\rm cm} \\  \d  y_{\rm cm}  \\ \d  z_{\rm cm}  \\  v_\rL \d t \end{array}\right). \label{eqxc}
\eeq
Equation (\ref{eq-3}), with substitutions from (\ref{18m}) and (\ref{eqxc}) yields four scalar equations which are precisely  the full set of equations in  (\ref{newtonpt1}) --- three scalar equations --- and (\ref{newtonptx}) --- one scalar equation. This result justifies the formulation of the matrix equation (\ref{eq-3}): the set of equations presented in subsection~\ref{ssec:newslaw}   for classical mechanics have been merged  into a single equation, hence equation (\ref{eq-3}) can be appropriately be called  `\emph{matrix centre-of-mass equation}' where ${\bf F}_{\rm ext}\, \, \d {\cal R}_{\, \rm cm}$, the external 4-vector impulse-pseudowork, is the counterpart of (\ref{fgty}). In general, the classical infinitesimal 4-vector impulse-pseudowork associated to a  force $\vec F$ is given by (compare with equations (\ref{fgty}) and  (\ref{explicit45}))
\beq
\delta  {\cal W}_{\rm ps}= \left(\begin{array}{c} v_\rL  \delta I_x  \\ v_\rL  \delta I_y \\  v_\rL  \delta I_z \\ \delta W _{\rm ps}  \end{array}\right) =
\left(\begin{array}{c} v_\rL F_x \d t  \\ v_\rL F_y  \d t \\  v_\rL F_z \d t \\ F_x \d x_{\rm cm} + F_y \d y_{\rm cm} + F_z \d z_{\rm cm}   \end{array}\right)\,
\eeq
(it is preferable to denote the infinitesimal work and impulse by $\delta$ because, in general, they are not exact differentials).

The matrix equation equivalent to (\ref{eq-3}) in reference S$'$, moving with velocity $V$ with respect to S along the $xx'$ axes, is readily obtained. We just have to apply the operator (\ref{upsilon}) to the left side of equation (\ref{eq-3}), i.e.
\beq
 \Upsilon \ \d  {\cal K}_{\, \rm cm} = {{ \Upsilon}}\  {\bf F}_{\rm ext}\  \Upsilon^{-1}\ \ \Upsilon\ \ \d  {\cal R}_{\, \rm cm}\, .
 \label{xxhgd}
\eeq
In general, any force remains invariant (as it should) after taking the limit $v_\rL \rightarrow  \infty$:
${\bf F}' = {{ \Upsilon}}\  {\bf F}_{\rm ext}\  \Upsilon^{-1} =
{\bf  F}
$.
Therefore, equation (\ref{xxhgd}) becomes
\beq
\d  {\cal K}'_{\, \rm cm} =   {\bf F}_{\, \rm ext}\   \d  {\cal R}'_{\, \rm cm}\, ,
\label{rtasd}
\eeq
so that the Principle of Relativity holds.

\subsection{ First law of thermodynamics in the  4-vector formalism}

The previous discussion applies to the centre-of-mass of a classical body in translation. However, in order to study the motion of an extensive deformable body one needs additional equations
to describe rotations (but we leave the discussion of the rotations to another publication) and
processes involving production
and destruction
of mechanical energy \cite{guemez13b}.
To address a complete description of these kind of processes
we also need the first law of thermodynamics
as discussed in section~\ref{sec:clasmeter}.

In the rest frame of the body, the fourth component of the internal energy includes  $Mv_\rL^2$ [see equation (\ref{eq-2})],
i.e. in the present formalism there is an energy associated with the object at rest.
However, we are mostly interested in ~kinetic or internal energy {\em variations}, therefore this constant rest energy does not play any role
and can be ignored.

In classical physics the internal energy of a body changes with no associated change of linear momentum.
Thus, the 4-vector internal energy variation is given by
\beq
\d {\cal U} =   \left( \begin{array}{c} 0 \\ 0 \\ 0 \\ \d U \end{array} \right) \, .
\eeq

In thermodynamics both work and heat contribute to the internal energy variation of a system.
For a general process on a macroscopic body, whose analysis needs to combine mechanics and thermodynamics \cite{jewett08v},
the four-vector formalism is still useful and we should consider the following
matrix equation that can be called `\emph{matrix energy equation}' (since it includes the conservation of the energy):
\beq
\d {\cal K}_{\rc \rmm} + \d {\cal U} = \sum_j \left( { {\bf F}_{{\rm ext}, j}}\ \d {\cal R}_j \right) \ + \ \delta {\cal Q}\, .
\label{totale}
\eeq
Here, the heat 4-vector has a structure similar to the internal energy \cite{kampen69}, i.e.
\beq
\delta {\cal Q} =   \left( \begin{array}{c} 0 \\ 0 \\ 0 \\  \delta Q \end{array} \right)\, .
\eeq
The first three components of  equation (\ref{totale}) are exactly the same as those in the matrix equation (\ref{eq-3}), and the same given by (\ref{newtonpt1}). However, the fourth component in (\ref{totale}) is precisely equation (\ref{totale1}) that formalizes the first law of thermodynamics. Since equation
(\ref{newtonpt1}) is equivalent to equation (\ref{newtonpt2}), we conclude that the redundancy contained in the matrix equation (\ref{eq-3}) does not exist in
(\ref{totale}). In fact, the latter includes, simultaneously, the Newton's second law (the first three components) and the first law of thermodynamics (the fourth component). Having these two fundamental laws expressed by a {\em single} equation clearly shows that both laws should be simultaneously valid, therefore they must be absolutely compatible. Hence, equation (\ref{totale}) is more general  than (\ref{eq-3}).

Each term in the sum on the right-hand side of equation~(\ref{totale})
is the 4-vector impulse-work associated with each {\em external} force ${\vec F}_j = (F_{x, \, j}, F_{y, \, j}, F_{z, \, j})$, that is
\beq
{{\bf  F}_{{\rm ext}, j}}\ \d {\cal R}_j = \delta {\cal W}_j =  \left(\begin{array}{cccc}0 & 0 & 0 &  F_{x ,\, j} \\0 & 0 & 0 & F_{y ,\, j} \\0 & 0 & 0 & F_{z ,\, j} \\  F_{x ,\, j} & F_{y ,\, j} &F_{z ,\, j} & 0\end{array}\right)\left(\begin{array}{c}  \d x_j\\ \d y_j\\ \d z_j \\  v_\rL  \d t \end{array}\right)\, ,
\eeq
where $\d {\cal R}_j$
is the 4-vector infinitesimal displacement of the  force $F_j$ itself (and not, anymore, the displacement of the centre-of-mass). Therefore, $\delta {\cal W}_j$ in the previous expression is associated to real work and not to pseudo-work.

Of course, for a body that behaves like an elementary particle (no rotation, no deformation, etc.) the centre-of-mass equation and the first law of thermodynamics provide the same information, and the same happens with the corresponding matrix equations. But when the body does not behave like an elementary particle, equation (\ref{totale}) provides
 additional information with respect to (\ref{eq-3}).

A final remark on the Galilean invariance of equation (\ref{totale}). The matrix equation (\ref{totale}) is Galilean invariant, similarly to equation (\ref{eq-3}):
\beq
\d {\cal K}'_{\rc \rmm} + \d {\cal U}' = \sum_j \left( { {\bf F}_{{\rm ext}, j}}\ \d {\cal R}'_j \right) \ + \ \delta {\cal Q}'\, .
\label{totaleprime}
\eeq
Due to the structure of the 4-vectors $\d {\cal U}$ and $\delta {\cal Q}$, these quantities are trivially Galilean invariants: after taking the limit
$v_\rL \rightarrow  \infty$, $\d {\cal U}'= \d {\cal U}$ and $\delta {\cal Q}'= \delta {\cal Q}$.

Summarizing, using the 4-vector formalism we established two matrix equations, the centre-of-mass equation and the energy equation (and the corresponding matrix transformation).
Both   equations are always valid and, therefore, should be compatible. However, the former contains some redundant information in its four components, whereas the latter provides additional  new information if the situation is within the scope of thermodynamics --- it includes simultaneously the Newton's second law and the first law of thermodynamics.
Of course, the second law of thermodynamics, which states that only processes compatible with a non-decrease of the entropy of the universe, $\Delta S_\rU \ge 0$, are allowed, should always be observed.

In the next section we discuss two examples to show the  4-vector formalism at work.

\section{The formalism at work}
\label{sec:tformwor}

In this section we discuss  two problems to illustrate with concrete examples how the formalism works.
Besides applying the formalism developed in the previous section in concrete situations, we also analyze the physical system from the point of view of the second law of thermodynamics.
These two examples are enough for illustration purposes but, of course, the formalism is applicable in general, without restrictions.

\subsection{Vertical jump}
\begin{figure}[hbt]
\begin{center}
\hspace*{0.0cm}
\includegraphics[width=6.5cm]{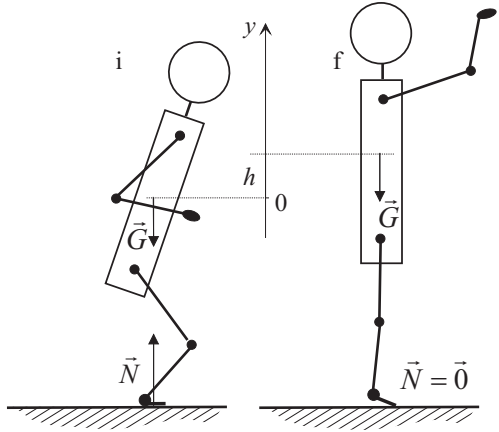}
\end{center}
\vspace*{-0.6cm}
\caption[]{\label{fig:jump} First phase of a vertical jump, during which there is still contact with the ground.}
\end{figure}

A person of mass $M$ jumps vertically as  figure~\ref{fig:jump} shows. We are interested in describing the motion while there is still contact with the ground.
We assume the following simplifying assumptions: the centre-of-mass motion is vertical, and the normal force, $\vec N$, is constant. The first assumption leads to a simpler description of the motion, which is then in 1+1 dimensions. The other assumption --- treating the contact force as an average constant force --- just makes easier the integration of the matrix equation (\ref{totale}).

We describe the motion in the reference frame with the $y$ axis pointing upwards. The other spatial
coordinates $x$ and $z$ are not necessary  and the corresponding dimensions can simply be skipped from the matrices. Hence, the $4\times 4$ matrices describing ${\vec N} = (N, 0, 0)$ and $\vec G$  reduce to  $2\times 2$ matrices, namely
\beq
{\cal N} = \left(\begin{array}{cc} 0&N\\N&0 \end{array}\right)\, , \ \ \ \ \ \  {\cal G} = \left(\begin{array}{cc} 0&-Mg\\-Mg&0 \end{array}\right) .
\eeq
The 4-vector displacements for the center-of-mass and for the forces are
\beq
{\cal R}_{\rm cm}  ={\cal R}_G= \left(\begin{array}{c} h \\ v_{\rL} t_0 \end{array}\right)\, , \ \ \
{\cal R}_N = \left(\begin{array}{c} 0 \\ v_{\rL} t_0 \end{array}\right)\,
\eeq
where $t_0$ is the duration of the action of the horizontal force upon the person and $h$ the corresponding spatial displacement of the center-of-mass.
Each of the above four-vectors now reduce to  two components: the first component is  space-like, and the second one is time-like.
The person is initially at rest and after time $t_0$, exactly  when the contact ceases, the centre-of-mass velocity is $v_{\rm cm}$.
The normal force, $\vec N$, is the reaction to the force exerted by the person on the ground and, to produce such a force, biochemical reactions should take place in the person's muscles.
 A chemical reaction $\xi$ produced inside the body may cause variations on its internal energy, $\Delta U_\xi$, volume, $\Delta V_\xi$, entropy, $\Delta S_\xi$, etc. The four-vector $\Delta \, {\cal U}_\xi$ associated with this internal energy variation is:
\beq
\Delta {\cal U}_\xi =   \left( \begin{array}{c}  0 \\ \Delta U_\xi\end{array} \right) \, .
\eeq
If a chemical reaction takes place at constant external pressure, $P$, in diathermic contact with a heat reservoir at temperature $T$, part of the internal energy is used for an expansion against the external pressure and part must be exchanged with the heat reservoir in order to ensure that the entropy of the universe does not decrease. Therefore, the four-vectors associated with the work, ${\cal W}_\xi$, and with the heat ${\cal Q}_\xi$ interaction are (again, we skip two space-like vanishing components):
\beq
{\cal W}_\xi =   \left( \begin{array}{c}  0 \\ - P \Delta V_\xi \end{array} \right) \, ; \ \ \
  {\cal Q}_\xi =   \left( \begin{array}{c} 0 \\  T \Delta S_\xi \end{array} \right)\, .
\eeq
When $\Delta V_\xi < 0$, the external pressure performs work on the system and when $\Delta S_\xi > 0$ the heat reservoir
increases the internal energy of the body. We assume that no other  energy is exchanged between the person and the environment.

The  matrix equation for this process --- equation (\ref{totale}) --- is:
\beq
\Delta {\cal K}_{\rc \rmm} + \Delta\,  {\cal U}_\xi  = {\cal W}_{G} + {\cal W}_{N}+ {\cal W}_{\xi} + {\cal Q}_\xi\, ,
\label{lk987}
\eeq
(${\cal W}_{G}$ and  ${\cal W}_{N}$ are the impulse-work 4-vectors related to the weight and the normal force)  or, explicitly,
 \beq
\left(\begin{array}{c}  v_\rL M v_{\rm cm}  \\  {1\over 2} M  v_{\rm cm}^2   \end{array}\right)+
\left(\begin{array}{c}  0 \\    \Delta U_\xi   \end{array} \right) =
\left(\begin{array}{c}  v_\rL (N-Mg) t_0  \\ -M\, g\, h -P\Delta V_\xi  \end{array}\right) +
  \left(\begin{array}{c}  0 \\    T\Delta S_\xi \end{array} \right)
\eeq
This matrix equation is equivalent to
\beq
\left\{
\begin{array}{rcl}
v_\rL \Big[ M v_{\rm cm} &=&  (N-Mg)  t_0 \Big] \, , \vspace*{0.3cm}\\
{1\over 2} M v_{\rm cm}^2 + \Delta U_\xi &=& -Mgh -P \Delta V_\xi + T\Delta S_\xi\, .
\end{array}
\right.
\label{eq:fltboys}
\eeq
The first of these equations is equivalent to the (centre-of-mass) equation
\beq
{1\over 2} M v_{\rm cm}^2 + Mgh= N h
\label{yudfg}
\eeq
i.e. the increase of mechanical energy in the process equals the pseudo-work of the normal force.
The second equation in equation~(\ref{eq:fltboys}) can also be expressed as
\beq
{1\over 2} M v_{\rm cm}^2 +Mgh = - \Delta G_\xi\, ,
\label{anter}
\eeq
where $\Delta G_\xi = \Delta U_\xi +P \Delta V_\xi - T\Delta S_\xi$
is the Gibbs free energy variation, which is symmetric of the pseudo-work associated to $\vec N$. Equation (\ref{anter})
shows that the person increases its center-of-mass kinetic and potential energy thanks to internal biochemical reactions \cite{atkins10}.
The variation of the Gibbs free energy is the maximum useful work that can be obtained from the reactions in the  person's muscles biochemical reactions: $W_{\rm max} = -  \Delta G_\xi$ and this is also equal to $Nh$. The normal force does not do any work but it intermediates an energy transformation.

The initial phase of  this idealized  jump is a process that implies no entropy increase of the universe and, therefore, it is reversible. In fact, the mechanical energy acquired by the system can be completely used, at least in principle,  to increase by $-\Delta G_\xi$ the free energy of any chemical reaction.

Let us now briefly analyze, from the mechanical and thermodynamical points of view,
the opposite process, i.e. the deceleration process of the body when  the person falls down,  touches the ground and stops. When the foot first comes in contact with the ground (the new initial time), the centre-of-mass is moving downwards with velocity
$-v_{\rm cm}$  and it stops at instant $t'_0$ after a vertical displacement  $h'$ at a supposedly constant acceleration. The (constant) reaction force is now $N'$, pointing upwards, not necessarily equal to $N$.
Note that we are assuming the same value for the velocity of the centre-of-mass when the foot abandons the ground, in the ascending trajectory,  and when it comes in contact again with the ground, in the descending phase. This means that, in both instants, the centre-of-mass is at exactly the same level with respect to the ground.
The dynamics of the descendent body is now described by the following equivalent equations:
\beq
Mv_{\rm cm} = (N'-Mg)t'_0    \ \ \ \ {\rm or} \ \ \ \ \ -\left( {1\over 2} M v_{\rm cm}^2 + Mgh'\right) = -N' h'\, .
\label{fg7564}
\eeq
Combining these equations with the dynamical equations for the upward phase of the motion --- first equation in (\ref{eq:fltboys}) and equation (\ref{yudfg}) --- one immediately concludes that
in the ascendent and in the descendent phases the impulses and the pseudo-works of the external forces are the same:
\beq
I'=(N'-Mg)t'_0= (N-Mg)t_0 = I \ \ \ \ {\rm and} \ \ \ \    W'_{\rm ps}=(N'-Mg)h'= (N-Mg)h = W_{\rm ps}\, .
\eeq

Regarding thermodynamics, the situation is not symmetric with respect to the ascendent phase. Now there is no internal energy variation (or thermodynamical work and heat) associated with chemical reactions.
The `time-like' component in the matrix equation  (\ref{totale}), after integration, is now simply given by
\beq
-{1\over 2} M v_{\rm cm}^2 = Mgh' + Q'
\eeq
where $Q'$ denotes the heat exchanged with the surroundings (considered a heat reservoir at temperature $T$). From this equation and from the second equation in (\ref{fg7564})
one obtains \hbox{$Q'=-N'h'<0$}. Again, the normal force doesn't do any work --- it rather intermediates the transformation of mechanical energy into heat. Being a negative quantity, this  is heat transferred from the body to the surrounding.
Considering the overall process --- ascendent and descendent phases ---, and  assuming that the thermodynamical initial and final states of the body are the same, one concludes that the entropy of the universe
increases by $\Delta S_{\rm U} = {N'h'\over T}>0$, due to the heat transfer to the heat reservoir. We note that there is a clear thermodynamical asymmetry between the upward phase and the downward one. The description of the part of the motion where the person is only subjected to its own weight (and assuming negligible air resistance forces) is  trivial and  irrelevant for the thermodynamical description of the overall process.

\subsection{A block pulled on an horizontal plane}

\begin{figure}[hbt]
\begin{center}
\hspace*{0.0cm}
\includegraphics[width=10cm]{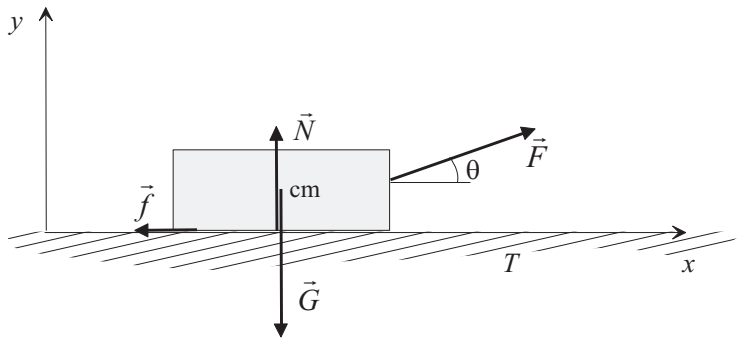}
\end{center}
\vspace*{-0.6cm}
\caption[]{\label{fig:block} Moving block on a horizontal surface with friction, under the action of a pulling force.}
\end{figure}

The next example consists of a rigid  block of mass $M$  pulled by a force $\vec F$, as  figure~\ref{fig:block} shows. The process is described in the reference frame represented there.

There are several external forces which are schematically represented in figure \ref{fig:block}: the pulling constant force, $\vec F$ (making the angle $\theta$ with the horizontal direction), the weight, $\vec G$, the normal reaction, $\vec N$, and the kinetic friction force, $\vec f$. This force is assumed to be constant, of the form $\vec f= \mu_{\rm k} N$, where $\mu_{\rm k}$
is the coefficient of kinetic friction.
We should note that  ${\vec N} + {\vec f}={\vec R}$ is a single force, resulting from the contact interaction between the block and the ground. However, it is usual to look at this force as the sum of ${\vec N}$ and ${\vec f}$ to better underline the role played by  each of these orthogonal forces in the dynamics of the body.

According to equation
(\ref{18m}), the matrix form of these forces, suitable for our formalism, is readily written down. Since the problem is 2+1 dimensional, we can skip one line and one column, corresponding to the space coordinate $z$. Therefore, the forces are expressed by:
\vspace*{2mm}
\beq
\!\, {\bf F}\!\! = \!\! \left(\!\! \begin{array}{ccc}0 & 0  &  F \cos \theta \\0  & 0 & F \sin \theta  \\  F \cos \theta & F \sin \theta & 0\end{array} \!\! \right) \, \,
{\bf G}\!\!  =\!\! \left(\!\! \begin{array}{ccc}0 & 0  &  0 \\0  & 0 & Mg  \\ 0 & Mg & 0\end{array}\!\! \right) \, \,
{\bf N}\!\! =\!\! \left(\!\! \begin{array}{ccc}0 & 0  &  0 \\0  & 0 & N  \\ 0 & N & 0\end{array}\!\! \right)\, \,
{\bf f} \!\!=\!\! \left(\!\! \begin{array}{ccc}0 & 0  & -f \\0  & 0 & 0  \\  -f  & 0 & 0\end{array}\!\! \right)\! .
\eeq

\noindent The magnitudes of all these forces are constant, hence the integration of equation (\ref{totale}) is straightforward. In particular, the work associated with each force is simply given by the dot product of the force and its own displacement.

We denote the space displacement of the centre-of-mass by $\Delta x_{\rm cm}$, which occurs in the time interval $t_0$. Therefore, the centre-of-mass 4-displacement is represented, in the formalism,
by the following column matrix now with just three lines:
\beq
{\cal R}_{\rm cm} = \left(\begin{array}{c}  \Delta x_{\rm cm} \\  0 \\  v_\rL  t_0  \end{array}\right)\, .
\eeq
Clearly, this is the displacement of the pulling force and of the weight:
${\cal R}_F = {\cal R}_G = {\cal R}_{\rm cm}$. Regarding the contact force, $\vec R$,  its origin ultimately lies in the countless local microscopic interactions of electromagnetic origin between the body and the ground. The application point of the macroscopic force $\vec R$ is such that the total torque on the block vanishes. For instance, if the pulling force direction includes the centre-of-mass, the reaction force direction should pass through the centre-of-mass as well. Contrary to $\vec F$ and $\vec G$, that are always present,  $\vec R$ is a force that constantly `appears' and immediately `disappears' as the body moves along the horizontal surface. In fact, we may think on a manyfold of forces, each of which acts at every location of the body. Therefore, it is tricky --- just to say the minimum --- to figure out what the displacement of {\em the} force $\vec R$ is.  If we think about the decomposition of $\vec R$ into $\vec N$ and $\vec f$ the work associated to the former is zero anyway, because it is always perpendicular, by definition, to the displacement whatever it is. Regarding the impulse of $\vec N$, it is neither dependent on its application point nor on its displacement. On the other hand, the space displacement associated to the kinetic friction force is more awkward. Some authors, exploring microscopic models for the friction based on electromagnetic molecular interactions, claim that the displacement of the kinetic friction force is half of the displacement of the block \cite{sherwood84}.
Others come to different conclusions \cite{poon06}.
Jewett  and  Serway argue that the work of the friction force is simply not calculable from the expression that defines work \cite{serway}.
As we said above, what happens is that the body experiences {\em different} friction forces, all with the same magnitude and direction but exerted at {\em different} locations.
Therefore, identifying the product of the constant friction force by the centre-of-mass displacement with the work done by that  force is questionable.

 Fortunately, the mentioned situation is not problematic if we take into account, on the right-hand side of equation (\ref{totale1}), the energetic interaction between the block and the surrounding originated by the contact force. In fact, it really does not matter whether that energy transfer is accounted as work (better to say {\em dissipative} work) or as heat. Actually, dissipative work  goes immediately into heat so we can treat thermodynamically any dissipative work and heat. To do so, and since the displacement of the force is not a well-defined concept, we may simply adopt the point of view that a kinetic friction force, similarly to a static friction force, does not do any work, operationally by assuming that its space displacement is zero. (Of course, there is pseudo-work  associated to the frictional force that is simply its magnitude times the centre-of-mass displacement, $W_{\rm ps}=-f \Delta x_{\rm cm}$.) The bottom line is that we may assume for the 4-displacement of $\vec f$ the following column matrix:
\beq
{\cal R}_{f} = \left(\begin{array}{c}  0 \\  0 \\  v_\rL  t_0  \end{array}\right).
\label{cmta}
\eeq
For the normal force we may also take the same 4-vector displacement, ${\cal R}_{N}= {\cal R}_{f}$, but it is irrelevant whether one takes this or ${\cal R}_{N}= {\cal R}_{\rm cm}$ because the space-like components of the 4-impulse-work will be always the same, and the time-like component will  also always be the same (zero).
We shall see that this point of view, formalized in the choice (\ref{cmta}), does  not affect the mechanical description of the block's motion and it does provide a very clean framework to discuss the thermodynamics of the process.

The 4-vector equation (\ref{totale}), whose integral form is $\Delta {\cal K}_{\rm cm} + \Delta {\cal U} = \sum_j {\cal W}_j+ {\cal Q}$, leads explicitly to
\beq
\left(\begin{array}{c}  v_\rL M v_{\rm cm} \\  0 \\  {1\over 2} M  v_{\rm cm}^2   \end{array}\right)+
\left(\begin{array}{c}  0 \\  0 \\  \Delta U   \end{array} \right) =
\left(\begin{array}{c}  v_\rL (F \cos \theta - f) t_0  \\  v_\rL (N+ F\sin \theta - Mg) t_0  \\ F \cos \theta \Delta x_{\rm cm}  \end{array}\right) +
  \left(\begin{array}{c}  0 \\  0 \\  Q  \end{array} \right)
  \label{asdior}
\eeq
where $v_{\rm cm}$ is the centre-of-mass velocity of the body, initially at rest, at instant $t_0$. The quantities $\Delta U$ and $Q$ are, respectively, the variation of the block internal energy and the heat transfer in the process.  Just for the sake of illustration, we may assume a simple model for the internal energy variation of the block, taking it as $\Delta U = M c (T_{\rm f}- T)$, where $c$ is the constant specific heat, $T_\rf$ is the final temperature and $T$ the initial temperature --- also the temperature of the ground that is supposed to be a heat reservoir. The previous matrix equation leads to
\beq
\left\{
\begin{array}{l}  v_\rL \Big[ \,  M\, v_{\rm cm} = (F\cos \theta - f ) \,  t_0\,  \Big] \vspace*{0.2cm}\\
v_\rL \Big[ \, 0 = (N+F\sin \theta - Mg) \,  t_0  \, \Big]\vspace*{0.2cm} \\
{1\over 2} M \, v_{\rm cm}^2 + M c (T_{\rm f}- T) =  F \cos \theta \, \Delta  x_{\rm cm} + Q \end{array}\right.
\label{xcv}
\eeq

The first two equations express the Newton's second law, whereas the third one expresses the conservation of energy. The second equation simply gives the value of $N$ and the  first equation is still equivalent to the centre-of-mass equation, namely
 \beq
 \  M\, v_{\rm cm} = (F\cos \theta - f ) \,  t_0  \ \ \ \Leftrightarrow \ \ \ \  {1\over 2} M \, v_{\rm cm}^2 =(F\cos \theta - f ) \Delta  x_{\rm cm}
 \eeq
 where the right-hand side of the second equation now includes the pseudo-work associated to the kinetic friction force. Inserting this result into the third equation in (\ref{xcv}) one readily obtains the following energy balance equation
 \beq
 Q= - f  \Delta  x_{\rm cm} + M c (T_{\rm f}- T)
 \label{qfx}
 \eeq
with a clear physical meaning: due to friction, the internal energy  increases,
the block heats up and part of the energy that is not used in this heating process is simply transferred as heat to the heat reservoir.
Serway and Jewett clearly state that the increase in internal energy of the system is equal to the product of the friction force and the displacement of the block \cite{serjew94} but that is not work strictly speaking.
Of course, if the temperature variation is tiny, so that it might legitimately be ignored, there is ultimately a heat transfer $Q=-f \Delta  x_{\rm cm}<0$ to the ground. Should we have assumed a different perspective for the displacement of the friction force, such as being equal to the centre-of mass displacement, one would arrive at    $Q=  M c (T_{\rm f}- T)$. Then the question would be, where does the heat come from to increase the internal energy of the block's energy? Of course, there is an answer but it is not as clear as the answer provided  by  (\ref{qfx}) which can still be written in the form:
\beq
W_{\rm D}=f \Delta  x_{\rm cm} = M c (T_{\rm f}- T) -Q\,
\eeq
 i.e. the energy available from  friction, the dissipative work $W_{\rm D}$, partly goes to the body, increasing its temperature, the remaining part is transferred to the heat reservoir. Formally, we can just say that there is a direct conversion of dissipative work into heat and the bottom line is that {\em dissipative work} and {\em heat} can be treated on the same footing. By assuming a zero space displacement of the friction force,  in practice we are treating the dissipative work as heat from the outset, without any logical rupture in the physical description of the process.

It is interesting to analyze the problem also from the perspective of the second law of thermodynamics. The point of view that we have adopted for the energy transfer associated with friction is also valuable in the following discussion.
The non-negative variation of the entropy of the universe is, in the present case, the sum of the entropy variation of the reservoir and of the entropy variation of the block (the thermodynamic system): $\Delta S_{\rm U} = \Delta S_{\rm R}+\Delta S$. The first term is simply
$\Delta S_{\rm R}=-{Q\over T}$, where the minus sign accounts for the fact that $Q$ is being considered from the point of view of the block. From the reservoir's point of view that heat is symmetric (hence, positive).
To compute the entropy variation of the block we use an auxiliary reversible process leading to the same final state at temperature $T_\rf$, i.e.   $\Delta S = Mc \int {\d T \over T}$. Hence
\beq
\Delta S_{\rm U}= {f  \Delta  x_{\rm cm} - M c (T_{\rm f}- T) \over T} + Mc \log \left({T_\rf \over T} \right)\, .
\eeq
According to the second law, $\Delta S_{\rm U}\ge 0$, which establishes a condition for $T_\rf$. If we define $\Delta T=T_\rf-T$ and assume that $T_\rf \sim T$, the previous expression can still be written as
\beq
\Delta S_{\rm U}\simeq {f  \Delta  x_{\rm cm} \over T} - {Mc\, \xi^2 \over 2} \ge 0
\label{law2}
\eeq
where we have used the power series expansion $\log (1+\xi)\simeq \xi-{1\over2} x^2$, with $\xi={\Delta T \over T}$. From (\ref{law2}) we can immediately conclude that the temperature variation must obey the following inequality: $(\Delta T )^2\le 2 f \Delta x_{\rm cm} / (Mc)$.

If, after the mechanical process, we allow enough time for the block to get in thermal equilibrium with the heat reservoir, there is an additional heat transfer
$|Q'|=M c (T_{\rm f}- T)$ from the block to the heat reservoir. In this case the overall heat transfer to the ground is $f\Delta x_{\rm cm}$ and the block does not suffer any change in its thermodynamical state. Hence, the total increase of the universe entropy is $\Delta S'_{\rm U} = {f \Delta x_{\rm cm}\over T} \ge \Delta S_{\rm U} \ge 0$.

In summary, the `choice' (\ref{cmta}), is compatible with the mechanical description of the motion (actually, it has nothing to do with that), and allows for a very clear thermodynamical description. That is equivalent to assume a direct conversion of dissipative work into heat, which is allowed by the second law of thermodynamics. The energy balance expressed by the third equation (\ref{xcv}) can still be written as
\beq
F \cos \theta \, \Delta  x_{\rm cm} = {1\over 2} M \, v_{\rm cm}^2 + W_{\rm D}\, .
\eeq
The energy transferred to the system as work by the force $\vec F$,
partly serves to increase the kinetic energy of the centre-of-mass and, the  other part, is exhausted as dissipative work which, in turn, part goes to the body, increasing its internal energy, and the remaining part flows to the ground. Similarly to the role of the normal force in the example 5.1  that intermediates an energy transfer without doing work, in the present example we may interpret the role of the friction force as an intermediator of an energy transfer (kinetic energy into heat) without doing work.

If the external pulling force ceases, the body eventually stops and cools down. From the thermodynamical point of view the process undergone by the body is cyclic --- the body final state is the same as its initial state. There was a direct
conversion of work into heat, which is allowed by the second law (the opposite is not allowed!), with a natural universe entropy increase since the process is irreversible.

\subsubsection{Galilean invariance}

It is rather pedagogical to see explicitly how the matrix formalism automatically guarantees   the fulfilment of the principle of relativity.
Let us consider a reference frame S$'$ moving along
the common $xx'$ axes with velocity $V$. The (now $3 \times 3$) matrix transformation is given by (\ref{upsilon}).
If we apply this matrix equation to (\ref{asdior}), written in a way that the right-hand side is a vanishing vector, we have
\beq
\left(\begin{array}{ccc} 1 & 0 & -V/v_{\rm L} \\0 & 1 & 0  \\-V/v_{\rm L} & 0 &   1 \end{array}\right)
\left(\begin{array}{c}  v_\rL \left[  M v_{\rm cm}- (F \cos \theta - f) t_0  \right]  \\
 v_\rL (N+ F\sin \theta - Mg) t_0 \\
 {1\over 2} M  v_{\rm cm}^2   +  \Delta U   - F \cos \theta \Delta x_{\rm cm} -Q  \end{array} \right) =
\left(\begin{array}{c} 0  \\ 0  \\ 0  \end{array} \right)
\eeq
or, keeping only the leading terms in $v_\rL$,
\beq
\label{xcvp}
\left\{
\begin{array}{l}  v_\rL \Big[ \,  M\, v_{\rm cm} - (F\cos \theta - f ) \,  t_0\,  \Big]=0 \vspace*{0.2cm}\\
v_\rL  (N+F\sin \theta - Mg) \,  t_0   =0 \vspace*{0.2cm} \\
-V \Big[ \,  M\, v_{\rm cm} - (F\cos \theta - f ) \,  t_0\,  \Big] + {1\over 2} M \, v_{\rm cm}^2 + M c (T_{\rm f}- T) -  F \cos \theta \, \Delta  x_{\rm cm} - Q =0 \end{array}\, . \right.
\eeq
Inserting the first equation into the third one, one immediately arrives at (\ref{xcv}).
Note that the formalism naturally introduces a `product force-displacement' $f V t_0$ required to ensure compliance  with the principle of relativity.
Of course, we could explicitly compute the 4-vectors of the equation
$\Delta {\cal K}'_{\rm cm} + \Delta {\cal U}' = \sum_j {\cal W}'_j+ {\cal Q}'$ in reference frame S$'$. The result would be the set of equations (\ref{xcvp}). This explicitly illustrates that the amount of physical information is always the same, independent of the chosen inertial reference frame.

\section{Conclusions}
\label{sec:conclu}

We have developed  a matrix formalism for classical  mechanics useful to deal, among others, with problems involving production or destruction  of mechanical energy.
Actually we put forward two matrix equations --- the  centre-of-mass and the energy equation ---, the former embodying Newton's second law and the latter also including the first law of thermodynamics.
Both equations are applicable to all situations, but we emphasize that there is additional information that can be obtained from the second with respect to the first one.

The classical formalism developed in this paper, inspired by the special relativity, automatically ensures the observance of the Principle of Relativity:
the two fundamental 4-vector  equations keep the same form in any inertial reference frame and,  irrespective of the frame, they provide {\em exactly} the same amount of information.
In  the resolution of physical problems we use most effective (simpler calculations) reference frame. However, it is important to realize that
any other inertial frame could be used, and the information would be always the same. We have done this explicitly for one example, taking advantage of the appropriateness of our formalism to this purpose.
Through the transformation matrix between inertial frames, transformations of 4-vectors are readily obtained. Regarding the $4\times 4$ matrices associated with the forces, they are invariant under these transformations.

In order to stress the ideas and keep the formalism and the discussion at the simplest level, we deliberately skipped any reference to rotations.
However, our  4-vector formalism can be extended to include rotations, a problem to be treated specifically in a forthcoming publication.
With respect to the present formalism, there will be a complementary one for rotations also applicable to  conservative or dissipative phenomena.

When we presented some examples to which the formalism was applied, we also discussed an interesting issue related to the role of the friction force, in particular to its work. For the mechanical description of the motion, the pertinent quantity is the pseudo-work, so any hypothesis about the kinetic friction force application point displacement does not matter at all.  However, for the discussion of energetic issues, treating the dissipative work associated to that force, i.e., the  associated energy transfer, as heat, leads us to a consistent and clear description, also from the point of view of the second law of thermodynamics: under that assumption we could even compute the universe entropy increase in a straightforward way.

The present 4-vector formalism was basically inspired by relativistic mechanics and shaped to deal with classical  mechanics problems.
We do not claim that the ideas expressed in this paper are directly applicable in the classroom, i.e. the formalism is not meant to provide a framework to teach
classical mechanics in introductory General Physics courses. However, the bridges it is able to establish between classical and relativistic physics, while allowing for a broader view on classical mechanics problems that includes simultaneously dynamics and the first law of thermodynamics, is certainly interesting both to teachers still in training or, as the authors, teachers with many years of experience. The formalism may also be appreciated by students already acquainted with classical mechanics, relativity and thermodynamics.

\end{document}